\documentclass[conference]{IEEEtran}

\usepackage{graphicx}
\usepackage{epstopdf}
\usepackage{amssymb,amsmath,amsthm}
\usepackage{xifthen}
\usepackage[serbian,english]{babel}
\usepackage[T1]{fontenc}
\usepackage{xcolor}
\usepackage[caption=false, font=footnotesize]{subfig}
\usepackage{array}

\newcommand{\ing}[1]{\mathsf{#1}}
\newcommand{\Rn}[1]{\ifthenelse{\equal{#1}{}}{\mathbb{R}}{\mathbb{R}^{\ing{#1}}}}
\newcommand{\Cn}[1]{\mathbb{C}^{\ing{#1}}}
\newcommand{\Rset}[2]{\ifthenelse{\equal{#2}{1}}{\in \Rn{\ing{#1}}}{\in \Rn{\ing{#1} \times \ing{#2}}}}
\newcommand{\Cset}[2]{\ifthenelse{\equal{#2}{1}}{\in \Cn{\ing{#1}}}{\in \Cn{\ing{#1} \times \ing{#2}}}}
\newcommand{\vect}[1]{\boldsymbol{\mathbf{\MakeLowercase{#1}}}}
\newcommand{\mtrx}[1]{\boldsymbol{\mathbf{\MakeUppercase{#1}}}}

\newcommand{\norm}[2]{\|#1\|_{#2}}

\newcommand{\transp}[1]{#1^{\mathsf{T}}}

\newcommand{\ind}[2]{\ifthenelse{\equal{#2}{}}{\chi_{#1}}{\chi_{#1}\left( #2 \right)}}

\DeclareMathOperator*{\minim}{minimize\,}

\usepackage{graphicx}
\graphicspath{{images/pdf/}}
\usepackage{amsmath,amssymb}
\usepackage{algorithm}
\usepackage{algorithmic}
\usepackage{flushend}

\begin{document}
\title{A Comparative Study of Multilateration Methods for Single-Source Localization in Distributed Audio}
\date{}

\author{\IEEEauthorblockN{Sr\dj{}an Kiti\'c, Cl\'ement Gaultier, Gr\'egory Pallone}
\IEEEauthorblockA{Orange Labs \\Cesson-S\'evign\'e, France \\ {srdan.kitic, clement.gaultier, gregory.pallone}@orange.com}}
\maketitle

\begin{abstract}
In this article we analyze the state-of-the-art in multilateration - the family of localization methods enabled by the range difference observations. These methods are computationally efficient, signal-independent, and flexible with regards to the number of sensing nodes and their spatial arrangement. However, the multilateration problem does not admit a closed-form solution in the general case, and the localization performance is conditioned on the accuracy of range difference estimates. For that reason, we consider a simplified use case where multiple distributed microphones capture the signal coming from a near field sound source, and discuss their robustness to the estimation errors. In addition to surveying the relevant bibliography, we present the results of a small-scale benchmark of few ``mainstream'' multilateration algorithms, based on an in-house Room Impulse Response dataset.
\end{abstract}

\renewcommand{\figurename}{Fig.}

\section{Introduction}

As the audio technologies incorporating distributed acoustic sensing -- like Internet Of Audio Things \cite{turchet2020internet} -- gain momentum, the questions regarding efficient exploitation of such acquired data naturally arise. A valuable information that could be provided by these networks is the location of the sound source, which can be a daunting task in the adverse acoustic conditions, namely in the presence of noise and reverberation. On the flipside, localization is usually only a pre-processing block of a larger processing chain (\emph{e.g.} in the case of location-guided separation and enhancement \cite{vincent2018audio}), thus its computational efficiency is of uttermost importance.



\begin{figure}
    \centering
    \includegraphics[clip, trim=0mm 15mm 0mm 15mm, width=0.7\linewidth]{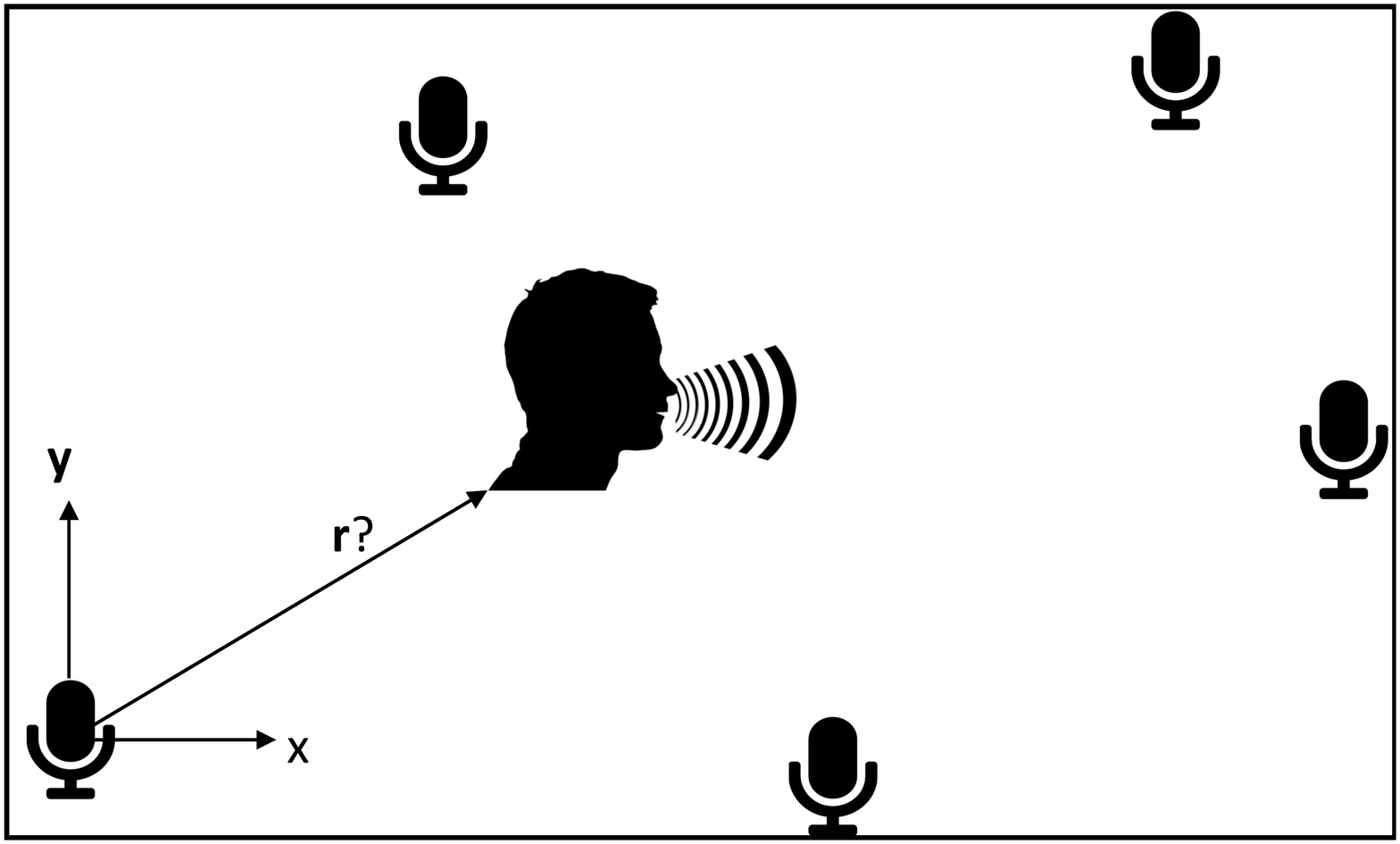}
    \caption{Source localization with 5 single-channel microphones.}
    \label{fig:room}
\end{figure}

In this work we consider a specific \emph{distributed audio context}, where we assume a sensor network composed of distributed single-channel microphones with potentially different gains (Fig.\ref{fig:room}). Such a network could be seen as one large scale microphone array - note that this is markedly different from a network whose nodes are (compact) microphone arrays themselves, as detailed in the following paragraph. The array geometry is assumed known in advance, and the microphones are already synchronized/syntonized \cite{joubaud2021eusipco}. Lastly, we assume the presence of a single sound source and a direct path (line-of-sight) between the source and microphones. The latter assumption is essential for most of the traditional sound source localization methods to work, in order to avoid an extremely challenging ``hearing behind walls'' problem \cite{kitic2014hearing}. 

%

Albeit deceivingly simple, the considered scenario imposes several technical constraints. First, the absence of compact arrays prevents the node-level Direction-of-Arrival (DOA) estimation. Second, no knowledge of the source emission time prohibits the Time-of-Flight (TOF) estimation. Third, large array size implies significant spatial aliasing, which, along with the relatively small number of microphones, seriously degrades performance of beamforming-based techniques, at least in narrowband \cite{dmochowski2008spatial}. The approaches based on \emph{distributed} beamforming, \emph{e.g.} \cite{valenzise2008resource,yao1998blind,hummes2011robust}, could still be appealing if they operate in the wideband regime: unfortunately, the literature on wideband beamforming by distributed mono microphones is somewhat scarce. Another major downside of beamforming-based localization is generally high computational cost, although various attempts have been made in order to reduce its complexity, \emph{e.g.} \cite{cobos2010modified}. Finally, the fact that the number of sensors and their spatial arrangement can vary precludes the use of contemporary learning-based localization methods (\emph{e.g.} \cite{vera2018towards, chakrabarty2019multi, perotin2019crnn}), which have shown remarkable performance in more restricted use cases. 

\begin{figure}
    \centering
    \includegraphics[clip, trim=0mm 0mm 0mm 0mm, width=1\linewidth]{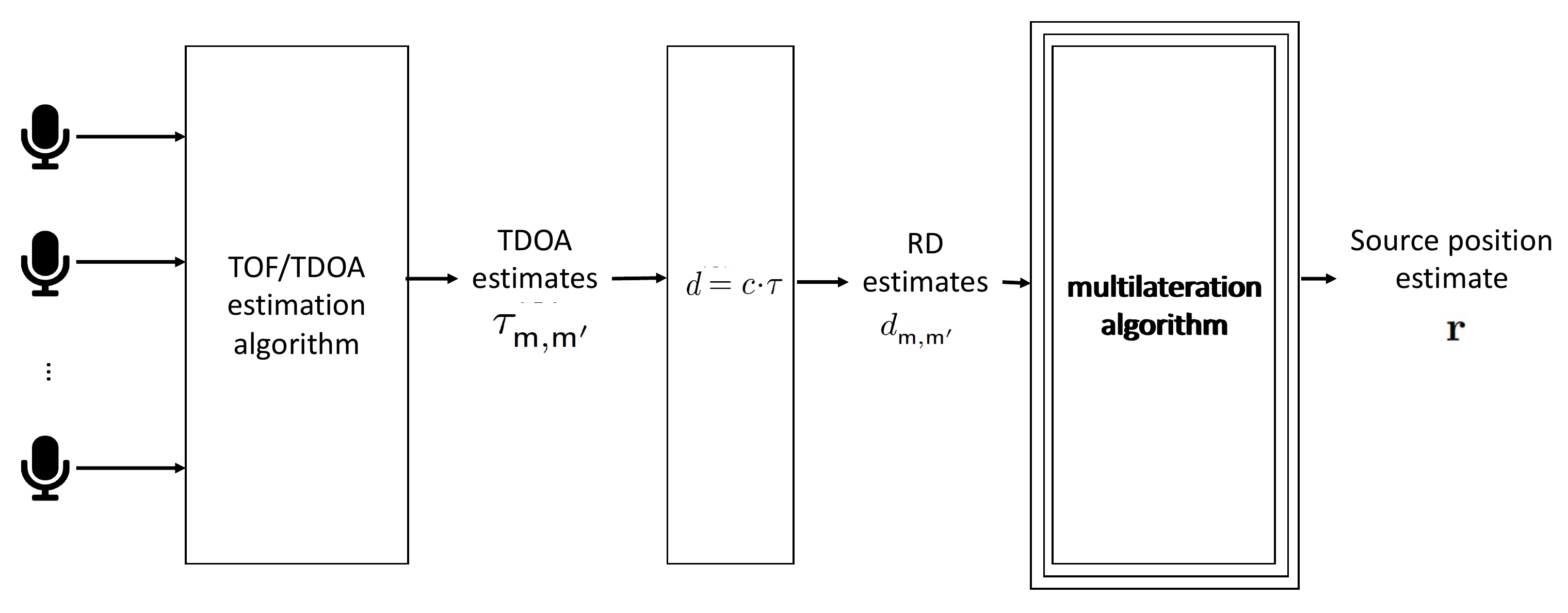} 
    \caption{Standard multilateration processing chain (notations are defined in section \ref{ProblemStatement}).}
    \label{fig:path}
\end{figure}

Under these constraints, the pairwise Time Difference Of Arrival (TDOA) features emerge as a viable choice for sound source localization (rectifying the need for adequate synchronization, since the TDOA estimation is known to be sensitive to clock offsets and internal delays \cite{wang2013tdoa}). The corresponding sound source localization pipeline is presented in Fig.~\ref{fig:path}. First, TDOAs are estimated for each microphone pair, followed by their conversion to (pseudo) Range Differences (RD). These estimates are then processed by a \emph{multilateration} algorithm \cite{fresno2017survey}, which finally yields the source position. In this article we focus on the last part of the pipeline, \emph{i.e.} we discuss exclusively and thoroughly the localization algorithms belonging to the multilateration class, as opposed to related review papers \cite{fresno2017survey, cobos2017survey, cheng2012survey} that study localization methods in a broader sense. 

Simultaneous localization of multiple sound sources is of a great practical interest. From the perspective of a multilateration algorithm, as long as multiple \emph{sets} of RDs (corresponding to each sound source) are available, the localization of each source can be done independently of the rest. Therefore, discussing the single-source case only does not endure a loss of generality with regards to multilateration-based localization. However, we underline that the performance of multilateration methods is conditioned on the accuracy of TDOA/RD estimation, which is a difficult problem in its own right \cite{compagnoni2014comprehensive, chen2006time,blandin2012multi}. The problem becomes even more challenging in the presence of multiple overlapping sources \cite{lombard2010tdoa}, but it falls out of the scope of the present article. Indeed, as practitioners we are particularly interested in performance of multilateration algorithms using the pseudo RDs obtained from off-the-shelf TDOA estimators, such as Generalized Cross Correlation - PHAse Transform (GCC-PHAT) \cite{knapp1976generalized}.

\section{Problem statement}\label{ProblemStatement}

Distances between microphones are considered to be of the same order as the distances between the microphones and the source, hence -- with regards to the audible frequency range of speech -- we discuss the near field scenario.  The general formulation of the time-domain signal $y_{\ing{m}}(t)$, recorded at the $\ing{m}$\textsuperscript{th} microphone is given by the time-variant convolution:
\begin{equation}\label{eqConvGeneral}
    y_{\ing{m}}(t) = \int\limits_{0}^{t} a_{m} (t, \tau) x_{\ing{s}}(t - \tau) d\tau + n_{\ing{m}}(t),
\end{equation}
where $a_{m} (t, \tau)$ is the time-variant Room Impulse Response (RIR) filter, relating the $\ing{m}$\textsuperscript{th} microphone position $\vect{r}_{\ing{m}}$ with the source position $\vect{r}_{\ing{s}}$, $x_{\ing{s}}(t)$ is the source signal, and $n_{\ing{m}}(t)$ is the additive noise of the considered microphone. In \eqref{eqConvGeneral}, the microphone gains are absorbed by RIRs. In practice, various simplifications are commonly used instead of the general expression \eqref{eqConvGeneral}. Commonly, a free-field, time-invariant approximation is adopted, as follows \cite{gustafsson2003positioning}:
\begin{equation}
    y_{\ing{m}}(t) = a_m x_{\ing{s}}(t - \tau_{\ing{m}} ) + n_{\ing{m}}(t),
\end{equation}
where the offset $\tau_{\ing{m}}$ denotes the TOF value, which is proportional to the source-microphone distance.

The TDOA, corresponding to the difference in propagation delay between the microphones $\ing{m}$ and $\ing{m'}$, is defined as $\tau_{\ing{m},\ing{m'}} = \tau_{\ing{m'}} - \tau_{\ing{m}}$. In homogeneous propagation media, the TDOA values $\tau_{\ing{m},\ing{m}'}$ , directly translate into Range Differences (RD) $d_{\ing{m},\ing{m'}}$, given the sound speed $c$:
\begin{equation}\label{eqRDmulti}
    d_{\ing{m},\ing{m'}} := D_{\ing{m'}} - D_{\ing{m}} =  \norm{\vect{r}_{\ing{m'}} - \vect{r}_{\ing{s}}}{} - \norm{\vect{r}_{\ing{m}} - \vect{r}_{\ing{s}}}{} = c\cdot \tau_{\ing{m},\ing{m'}},
\end{equation}
where $D_{\ing{m}}$ denotes the source-microphone distance. The observation model~\eqref{eqRDmulti} defines the two-sheet hyperboloid with respect to $\vect{r}_{\ing{s}}$, with foci in $\vect{r}_{\ing{m}}$ and $\vect{r}_{\ing{m}'}$ \cite{robots2011, rodriguez2011theoretical}. Note that the RDs could be easily determined from the TOF measurements as well (since $D_{\ing{m}} = c\cdot \tau_{\ing{m}}$), provided that the emission time is known. Surprisingly, despite theoretical evidence, a simulation study in \cite{kaune2012accuracy} has revealed that the TOF and TDOA features seem to perform similarly in terms of localization accuracy.

Since the microphone signals are often corrupted by noise and reverberation, the TDOA measurements -- thereby RDs -- could be erroneous, which negatively affects the performance of localization algorithms. In the RD domain, such degradations are usually modeled by an additive noise term. Another cause of localization errors is the inexact knowledge of microphone positions. As shown in \cite{ho2007source}, the Cram\'er-Rao lower Bound (CRB) \cite{theodoridis2015machine} of the source location estimate increases rather quickly with the increase in the microphone position ``noise'' (fortunately, somewhat less fast in the near field, than in the far field setting). Finally, the localization accuracy also depends on the array geometry \cite{yang2006theoretical}, which is assumed arbitrary in our case.

In noiseless conditions, the number of linearly independent RD observations is equal to $\ing{M} - 1$, and all the remaining RDs could be calculated from such a set. However, in the presence of noise, considering the full set of observations (of size $\ing{M}(\ing{M}-1)/2$) may be useful for alleviating the noise-related degradation \cite{kaune2012accuracy, hahn1973optimum, schmidt1972new,velasco2016tdoa}. If only independent RDs are to be used, one microphone is chosen as a reference (the choice of which, under the same noise level,  does not affect the theoretical localization accuracy \cite{kaune2012accuracy}). The same microphone could be conveniently put at the coordinate origin, \emph{e.g.} $\vect{r}_{\ing{1}} = \vect{0}$, where $\vect{0}$ is the null vector. By denoting $\vect{r} := \vect{r}_{\ing{s}}$, from \eqref{eqRDmulti}, the non-redundant RDs are compactly given as
\begin{equation}\label{eqRD}
    d_{\ing{m'}} := d_{\ing{1},\ing{m'}} =  \norm{\vect{r}_{\ing{m'}} - \vect{r}}{} - \norm{\vect{r}}{}.
\end{equation}

Finally, given a minimal $\{ d_{\ing{m'}} | \; \ing{m'} \in [2,\ing{M}] \}$, or an extended $\{ d_{\ing{m},\ing{m'}} | \; (\ing{m}, \ing{m'})  \in [1,\ing{M}] \times [1,\ing{M}],  \ing{m} \neq \ing{m'}\}$ set of observations, along with microphone position $\left\{ \vect{r}_{\ing{m}} \right\}$, the goal now is to estimate source position $\vect{r}$. In the following sections, we discuss two major classes of such localization algorithms: maximum likelihood and least squares estimators.

\section{Maximum likelihood estimation}

Since the observations \eqref{eqRD} are non-linear, a statistically efficient estimate (\emph{i.e.} the one that attains CRB) may not be available. The common approach is to seek the maximum likelihood (ML) estimator instead.

Let $\hat{\vect{r}}$ and $\hat{d}_{\ing{m'}}(\hat{\vect{r}})$ denote the estimated source position, and the corresponding RD, respectively:
\begin{equation*}
    \hat{d}_{\ing{m'}}(\hat{\vect{r}}) = \norm{\vect{r}_{\ing{m'}} - \hat{\vect{r}}}{} - \norm{\hat{\vect{r}}}{} .
\end{equation*}

Under the hypothesis that the observation noise is Gaussian, the ML estimator is given as the minimizer of the negative log-likelihood \cite{chan1994simple, huang2001real}
\begin{equation}\label{eqML}
    \mathcal{L}(\vect{r}) = \transp{\left( \vect{d} - \hat{\vect{d}}(\hat{\vect{r}}) \right)} \mtrx{\Sigma}^{-1} \left( \vect{d} - \hat{\vect{d}}(\hat{\vect{r}}) \right),
\end{equation}
where ${\vect{d} = \transp{\left[ d_{\ing{2}} \; d_{\ing{3}} \hdots d_{\ing{M}} \right]}}$, ${\hat{\vect{d}}(\hat{\vect{r}}) = \transp{ \left[ \hat{d}_{\ing{2}}(\hat{\vect{r}}), \; \hat{d}_{\ing{3}}(\hat{\vect{r}}) \hdots \hat{d}_{\ing{M}}(\hat{\vect{r}}) \right]}}$, and $\mtrx{\Sigma}$ is the covariance matrix of the measurement noise.

Note, however that the Gaussian noise assumption for the RD measurements may not hold. For instance, the digital quantization alone can induce RD errors on the order of $2$ cm \cite{huang2008time}. Moreover, the ML estimators are proven to attain the CRB in the asymptotic regime, while the number of microphones (\emph{i.e.} the number of RDs) is often small. Therefore, non-statistical estimators, such as least squares, are often used in practice instead. Anyhow, in this section we discuss two families of methods proposed for the RD maximum likelihood estimation: the ones that aim at solving the non-convex problem \eqref{eqML} directly, and the ones based on convex relaxations. The former should not be confused for ``direct'' localization methods based on grid search, such as steered response power beamformer.

\subsection{Direct methods}

The problem~\eqref{eqML} is difficult to solve directly, due to nonlinear dependence of the RDs $\{\hat{d}_{\ing{m'}}(\hat{\vect{r}})\}$ on the position variable $\hat{\vect{r}}$. Early approaches, based on iterative schemes, such as linearized gradient descent and Levenberg-Marquardt algorithm \cite{foy1976position, ajdler2004acoustic}, suffer from sensitivity to initialization, increased computational complexity and ill-conditioning (though the latter could be improved using regularization techniques \cite{mantilla2015localization}). The method proposed in \cite{bishop2008optimal} exploits correlation among noises within different RD measurements, and defines a constrained ML cost function tackled by a Newton-like algorithm. According to simulation results, it is more robust to adverse localization geometries \cite{bishop2010optimality, yang2006theoretical} than \cite{foy1976position}, or the least squares methods \cite{smith1987closed,schmidt1972new,li2004least,gillette2008linear}. Another advantage of this method is the straightforward way to provide the initial estimate (however, as usual, global convergence cannot be guaranteed).

In the pioneering article \cite{chan1994simple}, the authors proposed a closed-form, two-stage approach, that approximates the solution of \eqref{eqML}. Firstly, the (weighted) unconstrained least-squares solution (to be explained in the next section) \label{fnLS} is computed, which is then improved by exploiting the relation between the estimates of the position vector and its magnitude. The minimal number of microphones, due to the unconstrained LS estimation is $5$ in three dimensions. It has been shown \cite{chan1994simple} that the method attains the CRB at high to moderate Signal-to-Noise-Ratios (SNRs). Unfortunately, it suffers from a nonlinear ``threshold effect'' - its performance quickly deteriorates at low SNRs. Instead, an approximate, but more stable version of this ML method has been proposed in \cite{chan2006exact}.
In addition, the estimator \cite{chan1994simple} comes with a large bias \cite{mantilla2015localization}, which cannot be reduced by increasing the amount of measurements. This bias has been theoretically evaluated and reduced in \cite{ho2012bias}.

The method proposed in \cite{wang2011importance} uses Monte Carlo importance sampling techniques \cite{theodoridis2015machine} to approximate the solution of the problem~\eqref{eqML}. As an initial point, it uses the estimate computed by a convex relaxation method. According to simulation experiments, its localization performance is on par with the convex method \cite{yang2009efficient}, but the computational complexity is much lower.

A very recent article \cite{larsson2019optimal} proposes the linearization approach that casts the original problem into an eigenvalue one, which can be solved optimally in closed form. Additionally, the authors propose an Iterative Reweighted Least Squares scheme that approximates the ML estimate for different noise distributions.

\subsection{Convex relaxations}

Another important line of work are the methods based on convex relaxations of ML estimation problems. In other words, the original problem is approximated by a convex one \cite{boyd2004convex}, which is usually far easier to solve. Two families of approaches dominate this field: methods based on semidefinite programming (SDP), and the ones relaxing the original task into a second-order cone optimization problem (SOCP). In the former, the non-convex quadratic problem \eqref{eqML} is first \emph{lifted} such that the non-convexity appears as a rank 1 constraint, which is then substituted by a positive semidefinite one \cite{ma2010semidefinite}. Lifting refers to a problem reformulation (by a suitable variable substitution), such that the original problem is redefined in a higher-dimensional space. The rationale behind lifting is that the new problem becomes easier to solve, despite being high dimensional (particularly, it leads to a SDP problem). On the other hand, solving the SDP optimization problems can be computationally expensive, and the SOCP framework has been proposed as a compromise between the approximation quality and computational complexity (\emph{cf.} \cite{kim2001second} and the references therein for technical details).

One of the first convex relaxation approaches for the RD localization is \cite{lui2008semidefinite}, based on SDP. The algorithm requires the knowledge of the microphone closest to the source, in order to ensure that all RDs (with that microphone as a reference) are positive. The article \cite{yang2009efficient} discusses three convex relaxation methods. The first one, based on SOCP relaxation is computationally efficient, but restricts the solution to the convex hull \cite{biswas2004semidefinite, boyd2004convex} of microphone positions. The other two SDP-based remove this restriction, but are somewhat more computationally demanding. In addition, one of these is the \emph{robust} version - it minimizes the worst-case error due to imprecise microphone locations. The latter requires tuning of several hyperparameters, among which is the variance of the microphone positioning error. All three versions are based on the white Gaussian noise model for the RD measurements, however, whithening could be applied in order to support the correlated noise case. However, the SDP solutions are not the final output of the algorithms, but are used to initialize nonlinear iterative scheme, such as \cite{foy1976position}.

Interestingly, a recent article \cite{qu2016efficient} has shown that the ideas of the direct approach \cite{chan1994simple} and the constrained least-squares approach could be mixed together. Moreover, the cost function can be cast to a convex problem, for which an interior-point method has been proposed. However, in practice, it is a compound algorithm which iteratively solves a sequence of convex problems in order to re-calculate a weighting matrix dependant on the estimated source position. The accuracy depends on the number of iterations, which, in turn, increases computational complexity. As for \cite{chan1994simple}, it requires $5$ microphones for the 3D localization.

\section{Least-squares estimation}\label{secLS}

Largely due to computational convenience, the least-squares (LS) estimation is often a preferred parameter estimation approach. It is noteworthy that all LS approaches optimize a somewhat ``artificial'' estimation objective, which can induce large errors in very low SNR conditions, when the measurement noise is not white, and/or for some adverse array geometries \cite{bishop2008optimal,sirola2010closed, ho2012bias}.

Three types of cost functions are discussed: hyperbolic, spherical and conic LS.

\subsection{Hyperbolic Least Squares}

The goal is to minimize the sum of squared distances $\epsilon_{\mathrm{h}}$ between the true and estimated RDs:
\begin{multline}\label{eqCostHyp}
    \epsilon_{\mathrm{h}}( \hat{\vect{r}} ) := \sum\limits_{\ing{m'}=2}^{\ing{M}} \left( d_{\ing{m'}} -  \norm{\vect{r}_{\ing{m'}} - \hat{\vect{r}}}{} + \norm{\hat{\vect{r}}}{} \right)^2  \\ =  \transp{\left( \vect{d} - \hat{\vect{d}}(\hat{\vect{r}}) \right)} \left( \vect{d} - \hat{\vect{d}}(\hat{\vect{r}}) \right) ,
\end{multline}
which is analogous to the ML estimation problem~\eqref{eqML} for $\mtrx{\Sigma} = \mtrx{I}$, with $\mtrx{I}$ being the identity matrix. Thus, in the case of \emph{white} Gaussian noise, the hyperbolic LS solution coincides with the ML solution. Otherwise, solving \eqref{eqCostHyp} comes down to finding the point $\hat{\vect{r}}$ whose cumulative distance $\sum d_{\ing{m'}}$ to all hyperboloids, defined in \eqref{eqRD}, is minimal.

However, the hyperbolic LS problem is also non-convex, and its global solution cannot be guaranteed. Instead, local minimizers are found by iterative procedures, such as (nonlinear) gradient descent or particle filtering \cite{torrieri1984statistical,gustafsson2003positioning}. Obviously, the quality of the output result of such algorithms depends on their initial estimates, the choice of which is usually not mathematical, but rather application-based.

\subsection{Spherical Least Squares}

By squaring the idealized RD measurement expression~\eqref{eqRD}, followed by some simple algebraic manipulations, we have
\begin{equation*}
    d_{\ing{m'}} \norm{\vect{r}}{} + \transp{\vect{r}_{\ing{m'}}}\vect{r} - \underbrace{\frac{1}{2} \left( \norm{\vect{r}_{\ing{m'}}}{}^2 - d_{\ing{m'}}^2 \right)}_{b_{\ing{m'}}} = 0.
\end{equation*}
The interest of this operation lies in decoupling of the position vector and its magnitude, which are to be replaced by their estimates $\hat{\vect{r}}$ and $\hat{D}:=\norm{\hat{\vect{r}}}{}$, respectively.

The goal now becomes driving the sum of left hand sides (for all microphones) to zero:
\begin{equation}\label{eqSphLS}
    \epsilon_{\mathrm{sp}} = \sum\limits_{\ing{m'}=2}^{\ing{M}} \left( d_{\ing{m'}} \hat{D} + \transp{\vect{r}_{\ing{m'}}}\hat{\vect{r}} - b_{\ing{m'}} \right)^2, \; \hat{D}^2=\norm{\hat{\vect{r}}}{}^2,
\end{equation}
which leads to the following (compactly written) constrained optimization problem \cite{beck2008exact}:
\begin{align}\label{eqConstrained}
    \minim_{\hat{\vect{c}}} & \norm{\mtrx{\Phi} \hat{\vect{c}} - \vect{b}}{}^2 \\
    \text{subject to} \; \transp{\hat{\vect{c}}} \left[
    \begin{smallmatrix}
        1 & \mtrx{0} \\
        \vect{0} & -\mtrx{I}
    \end{smallmatrix} \right] \hat{\vect{c}} & = 0 \; \text{and} \; \hat{c}_{(1)} \geq 0, \nonumber
\end{align}
where $\mtrx{\Phi} = \left[
    \begin{smallmatrix}
        d_{\ing{2}} & \transp{\vect{r}_{\ing{2}}} \\
        d_{\ing{3}} & \transp{\vect{r}_{\ing{3}}} \\
        \hdots  &   \hdots \\
        d_{\ing{M}} & \transp{\vect{r}_{\ing{M}}}
    \end{smallmatrix}    \right]$, $\hat{\vect{c}} = \left[ \begin{smallmatrix} \hat{D} \\ \hat{\vect{r}} \end{smallmatrix} \right]$, $\vect{b} = \left[ \begin{smallmatrix} b_{\ing{1}} \\ b_{\ing{2}} \\ \hdots \\ b_{\ing{M}} \end{smallmatrix} \right]$, and $\hat{c}_{(1)}$ denotes the first entry of the column vector $\hat{\vect{c}}$.

In the literature, the problem above is tackled as:

\paragraph{Unconstrained LS} by ignoring the constraints relating the position estimate $\hat{\vect{r}}$ and its magnitude $\hat{D}$, the problem~\eqref{eqSphLS} admits a closed-form solution $\hat{\vect{c}}^* = \left(\transp{\mtrx{\Phi}} \mtrx{\Phi} \right)^{-1} \transp{\mtrx{\Phi}} \vect{b}$. As pointed in \cite{stoica2006lecture,wei2008comments}, several well-known estimation algorithms \cite{smith1987closed,li2004least,gillette2008linear} actually yield the unconstrained LS estimate. The minimum of $\ing{M}=5$ microphones (\emph{i.e.} four RD measurements), in three dimensions, are required in order for $\left(\transp{\mtrx{\Phi}} \mtrx{\Phi} \right)^{-1}$ to be an invertible matrix.

\paragraph{Constrained LS} While the unconstrained LS is simple and computationally efficient, its estimate is known to have a large variance compared to the CRB \cite{wei2008comments}, hence the interest for solving the constrained problem. Unfortunately, \eqref{eqConstrained} is non-convex due to quadratic constraints. To directly incorporate the constraint(s), a Lagrangian-based iterative method has been proposed in \cite{huang2001real}, albeit without any performance guarantees.

Later, in their seminal paper \cite{beck2008exact}, Beck and Stoica provided a closed-form \emph{global} solution of the problem, and demonstrated that it gives orders of magnitude more accurate solution (at an increased computational cost) than the unconstrained LS estimator. Moreover, the results in \cite{wang2011importance} indicate that it is generally more accurate than the two-stage ML solution \cite{chan1994simple}.


\subsection{Conic Least Squares}

In \cite{schmidt1972new}, Schmidt has shown that (in two dimensions) the RDs of \emph{three} known microphones define the major axis of a general conic (a hyperbola, an elipse or a parabola), on which the corresponding microphones lie. In addition, the source is positioned on its focus. In three dimensions, this axis becomes a plane containing the source. The fourth (non-coplanar) microphone is needed to infer the source position $\vect{r}$, by calculating the intersection coordinates of three such planes (hence the name \emph{plane intersection method} in the literature \cite{smith1987closed}). Thus, the method attains the theoretical minimum for the required number of microphones for RD localization. Nevertheless, given the minimal number of measurements, multilateration often yields an ill-posed problem \cite{herath2013robust}. Thereby, in practice, more sensors are needed for obtaining a meaningful result.

To illustrate the approach, let one such triplet of microphones be described by $(\vect{r}_{\ing{1}}, \vect{r}_{\ing{2}}, \vect{r}_{\ing{3}})$, and $(D_{\ing{1}}, D_{\ing{2}}, D_{\ing{3}})$ -- their position vectors, and the distances to the source, respectively. For a pair $(\ing{i},\ing{j})$ of these microphones, we have the following expression for the product of the \emph{range sum} $\Sigma_{\ing{i,j}} = D_{\ing{i}} + D_{\ing{j}}$ and the \emph{range difference} $d_{\ing{i,j}} = D_{\ing{j}} - D_{\ing{i}}$:
\begin{equation}\label{eqSumDif}
    \Sigma_{\ing{i,j}}d_{\ing{i,j}} = \norm{\vect{r}_{\ing{j}}}{}^2 - \norm{\vect{r}_{\ing{i}}}{}^2  - 2 \transp{(\vect{r}_{\ing{j}} - \vect{r}_{\ing{i}} )} \vect{r}.
\end{equation}
Note here that the conic method uses a full set of RD observations, as opposed to the spherical least squares approach.

By rearranging the terms in \eqref{eqSumDif}, and having $d_{\ing{k,i}}=\Sigma_{\ing{i,j}} - \Sigma_{\ing{j,k}}$, the range sums can be eliminated. Eventually, this gives the aforementioned plane equation
\begin{multline}\label{eqPlane}
    \transp{\left( d_{\ing{2,3}} \vect{r}_{\ing{1}} + d_{\ing{3,1}} \vect{r}_{\ing{2}} + d_{\ing{1,2}} \vect{r}_{\ing{3}} \right)}\vect{r} \\ = \frac{1}{2} \left( d_{\ing{1,2}}d_{\ing{2,3}}d_{\ing{3,1}} + d_{\ing{2,3}} \norm{\vect{r}_{\ing{1}}}{}^2 + d_{\ing{3,1}} \norm{\vect{r}_{\ing{2}}}{}^2 + d_{\ing{1,2}} \norm{\vect{r}_{\ing{3}}}{}^2 \right).
\end{multline}
This is a linear equation of three unknowns, thus the exact solution is obtained when three triplets (\emph{i.e.} four non-coplanar microphones) are available. Browsing the literature, we found that exactly the same closed-form approach has been recently reinvented in the highly cited article \cite{bucher2002synthesizable}.

For $\ing{M}$ microphones, one ends up with $\ing{M}\choose{3}$ such equations (in 3D) - the classical LS solution is to stack them into a matrix form, and calculate the position $\vect{r}$ by applying the Moore-Penrose pseudoinverse. Let $A_{\ing{pqr}}$, $B_{\ing{pqr}}$, $C_{\ing{pqr}}$ and $F_{\ing{pqr}}$ denote the coefficients and the right hand side of the expression \eqref{eqPlane}, for the microphone triplet $\ing{m} \in \{\ing{p, q, r} \}$, respectively. For all such triplets, we have
\begin{equation}
    \underbrace{\left[
        \begin{matrix}
            A_{\ing{123}} & B_{\ing{123}} & C_{\ing{123}} \\
            \vdots        &     \vdots    &     \vdots  \\
            A_{\ing{pqr}} & B_{\ing{pqr}} & C_{\ing{pqr}} \\
            \vdots        &     \vdots    &     \vdots  \\
        \end{matrix}
    \right]}_{\mtrx{\Psi}}
    \vect{r} =
    \underbrace{\left[
        \begin{matrix}
            F_{\ing{123}} \\
            \vdots \\
            F_{\ing{pqr}} \\
            \vdots
        \end{matrix}
    \right]}_{\vect{\psi}},
\end{equation}
where
\begin{align*}
    A_{\ing{pqr}} &= d_{\ing{q,r}} r_{\ing{p}(1)} + d_{\ing{r,p}} r_{\ing{q}(1)} + d_{\ing{p,q}} r_{\ing{r}(1)}, \\
    B_{\ing{pqr}} &= d_{\ing{q,r}} r_{\ing{p}(2)} + d_{\ing{r,p}} r_{\ing{q}(2)} + d_{\ing{p,q}} r_{\ing{r}(2)}, \\
    C_{\ing{pqr}} &= d_{\ing{q,r}} r_{\ing{p}(3)} + d_{\ing{r,p}} r_{\ing{q}(3)} + d_{\ing{p,q}} r_{\ing{r}(3)} \; \text{and} \\
    F_{\ing{pqr}} &= \frac{1}{2} \left( d_{\ing{p,q}}d_{\ing{q,r}}d_{\ing{r,p}} + d_{\ing{q,r}} \norm{\vect{r}_{\ing{p}}}{}^2 + d_{\ing{r,p}} \norm{\vect{r}_{\ing{q}}}{}^2 + d_{\ing{p,q}} \norm{\vect{r}_{\ing{r}}}{}^2 \right),
\end{align*}
as in \eqref{eqPlane}. However, such LS solution is strongly influenced by the triplets having large $A_{\cdot}$, $B_{\cdot}$, $C_{\cdot}$ or $F_{\cdot}$ values. Instead, as proposed in \cite{schmidt1972new}, the matrix $\mtrx{\Psi}$ needs to be preprocessed prior to computing the pseudoinverse - its rows should be scaled by $1/\sqrt{A_{\cdot}^2 + B_{\cdot}^2 + C_{\cdot}^2}$, as well as the corresponding entry of the vector $\vect{\psi}$.

Likewise, the presence of noise in the RD measurements $d_{\ing{i,j}}$ could seriously degrade the localization accuracy. In that case, the observation model \eqref{eqRDmulti} contains an additive noise term, which varies accross different measurements, rendering them \emph{inconsistent}. This means that the intrinsic redundancy within RDs does not hold, \emph{e.g} $d_{\ing{i,k}} \neq d_{\ing{i,j}} + d_{\ing{j,k}}$. In the noiseless case, the vector $\vect{d}$ of concatenated RD measurements, lies in the range space of a simple first-order difference matrix \cite{schmidt1996least}, specified by \eqref{eqRDmulti} and the ordering of distances $D_{\ing{m}}$. Thus, the measurements could be preconditioned, by replacing them with the closest feasible RDs, in the LS sense. This is done by projecting the measured $\vect{d}$ onto the range space of a finite difference matrix, or, equivalently by the technique called ``TDOA averaging'' \cite{schmidt1996least}. While computationally efficient, the proposed preconditioning technique assumes Gaussian distribution of TDOA (or RD) estimation errors \cite{velasco2016tdoa}, which may produce suboptimal results.


\section{Speaker localization experiments}

To the best of our knowledge, to date, there is no comprehensive benchmark of multilateration algorithms in the context of distributed single-channel audio sensing. The aim of this section is to contribute by providing an empirical analysis of three algorithms representative of the least squares class. In particular, we conduct a small-scale benchmark of the conic least squares \cite{schmidt1972new}, unconstrained \cite{smith1987closed} and constrained \cite{beck2008exact} spherical least squares (referred to as conic, usrd-ls and srd-ls, respectively). We first present the setup, data and performance measures before proceeding to the benchmark results.

\begin{figure}[htbp]
    \centering
    \includegraphics[width=0.6\linewidth]{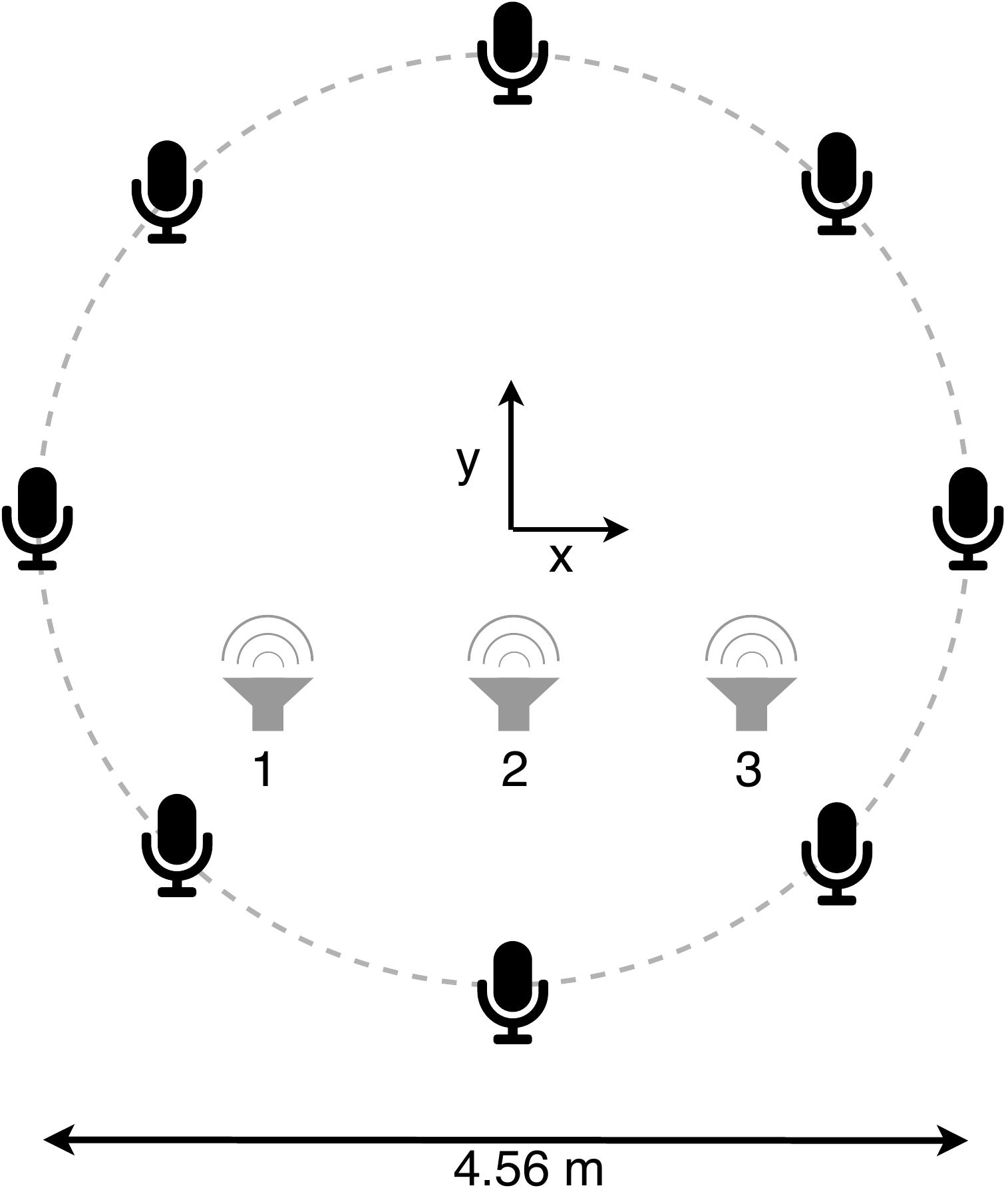}
    \caption{Experimental setup}
    \label{fig:ExpSetup}
\end{figure}

\begin{table}[h]
    \centering
    \caption{Source coordinates \& orientations}
    \label{tab:SourcePos}
    \begin{tabular}{l|c|c|c|}
    \cline{2-4}
                                          & Position 1     & Position 2     & Position 3     \\ \hline
    \multicolumn{1}{|l|}{x [cm]}      & -80            & 0              & 80             \\ \hline
    \multicolumn{1}{|l|}{y [cm]}      & -80            & -80            & -80            \\ \hline
    \multicolumn{1}{|l|}{z [cm]}      & 119            & 119            & 119            \\ \hline
    \multicolumn{1}{|l|}{Azimuth [$^{\circ}$]} & \{0, $\pm$90, 180\} & \{0, $\pm$90, 180\} & \{0, $\pm$90, 180\} \\ \hline
    \end{tabular}
\end{table}

The microphone signals are generated by convolving dry source signals with real RIRs measured in our audio lab whose $RT60=350~ms$. This approach allows to spare time and use different excitation signals afterwards. Eight omnidirectionnal DPA\textsuperscript{\textregistered} 4060 microphones, arranged in a circle of radius $2.28$m, surround the Genelec\textsuperscript{\textregistered} 1031A loudspeaker which serves as a source, as shown in \ref{fig:ExpSetup}. The microphones and the loudspeaker are approximately in the same horizontal plane (the difference in their $z$-coordinates is about $15$cm). The RIRs are retrieved using exponential sine sweep excitation signals ranging from 20 Hz to 20 kHz \cite{Farina2000}. During each recording, the loudspeaker is static, placed at one of the $3$ different positions within the circle (Table \ref{tab:SourcePos} details the source positions and azimuths). To account for the loudspeaker directivity, at each position, the loudspeaker is oriented in $4$ different directions, by rotating it around the $z$-axis in steps of $90^{\circ}$. The original sampling rate of the microphone recordings was $48$kHz (because that database will have other uses) and has been subsequently down-sampled to $16$kHz to match the signals described later, and all devices share the same clock.

For computing the TDOAs, we opted for the widely used GCC-PHAT  method. The signals are first segmented in $50\%$-overlapping frames of duration $0.064$s, and modified by Hanning window. The frame duration is chosen as the limit of the local (quasi-) stationarity of speech signals \cite{benesty2007springer}, to maximize the number of samples used for estimating correlations. Then, the TDOA estimation is performed for each frame pair, and a single output value is produced by median aggregation. For the aggregation, we either consider all frame-wise TDOAs (the ``no-VAD'', \emph{i.e.} no Voice Activity Detector setting), or we choose a subset of these using a simple energy-based criterion. For the latter, we evaluate the energy of each frame of the pair, and if neither has the energy that exceeds the half median energy of the sum of the two windowed representations, it is discarded (``VAD'' variant). At the end, the obtained TDOA matrix is either directly provided to a multilateration algorithm, or is further postprocessed by the TDOA averaging method \cite{schmidt1996least} (the ``denoised'' TDOA).

Concerning the algorithm-specific settings, we tested the conic LS with and without row-wise normalization, and spherical LS with two choices of the reference microphone. We refer to the choices as the ``max'' and ``min'', designating the microphones that produce the recordings of highest and lowest energy, respectively. This is motivated by the fact that, as the optimal choice of the reference microphone is not always straightforward, the corresponding non-redundant RD subset may contain more or less inaccuracies.

Regarding each algorithmic setting as a separate method, and the different TDOA variants as distinct features, the benchmark comprises $6$ multilateration methods, each applied to $4$ types of TDOA features. For each experiment, all possible subsets of $5$ out of $8$ microphones are selected. Note that for some loudspeaker orientations and microphone subsets the line-of-sight assumption is essentially violated (except at very low frequency bands). As the source signals, we use $6$ short audio excerpts from the TIMIT database \cite{garofolo1993darpa}, comprising $3$ male and $3$ female speakers (meaning that each experiment has been repeated $6$ times using a different excitation signal).

The multilateration performance is quantified as the position error in meters, between the ground truth and the estimated position, \emph{i.e.} $\norm{\vect{r} - \hat{\vect{r}}}{}$. In addition, we track the TDOA/RD estimation errors, in order to evaluate the robustness of a localization method to various levels of RD ``noise''. Therefore, we also store the average absolute errors of all RDs for the considered microphone subset (the ground truth RDs are easily determined from the microphone-source geometry).

\begin{figure*}[htbp]
    \centering
    \includegraphics[clip, trim=0mm 0mm 0mm 0mm, width=\textwidth]{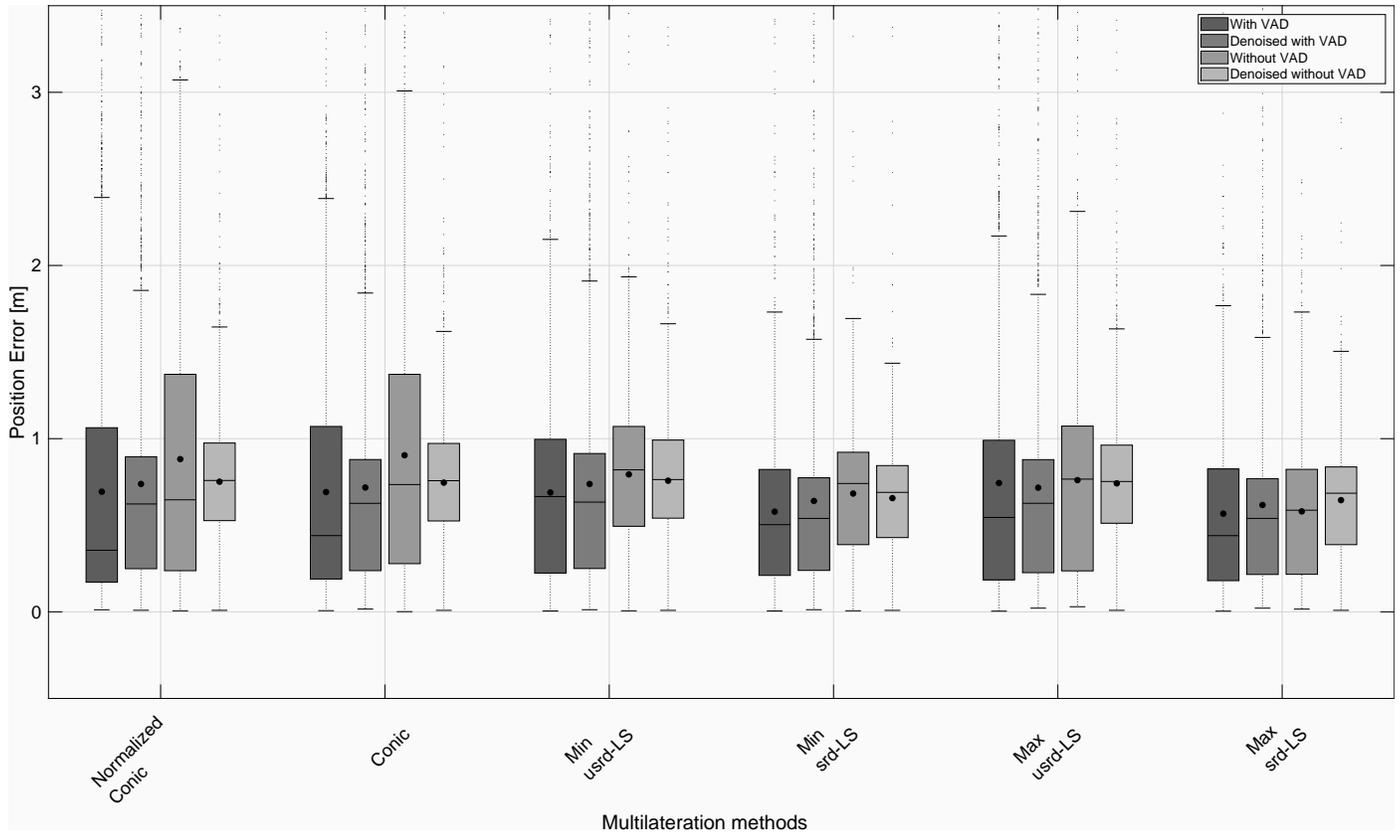}
    \caption{Source localization performance}
    \label{fig:LocaResults}
\end{figure*}

The overall results in the form of a box plot are shown on \ref{fig:LocaResults}, which we interpret below.

First, the RD denoising (by means of TDOA averaging) does not contribute to an increased localization accuracy. Although the variance of positioning errors is reduced for the denoised RDs, the median error is generally larger than for the non-processed RD features. The denoising operation actually redistributes the error across all RDs, thereby corrupting all the ’’clean'' RD entries as well. Contrary to the intuition, the conic LS method seem to be affected the most, probably due to the fact that it exploits all available RD observations, thereby accumulating most of the noise.

Second, our rudimentary VAD solution seems to be beneficial, despite the fact that the TIMIT source signals contain mostly uninterrupted speech. We attribute this to the selection of highly energetic frames, which are affected by noise and reverberation to a lesser degree.

\begin{figure*}[htbp]
    \centering
    \subfloat[Normalized Conic]{\includegraphics[width=0.33\textwidth]{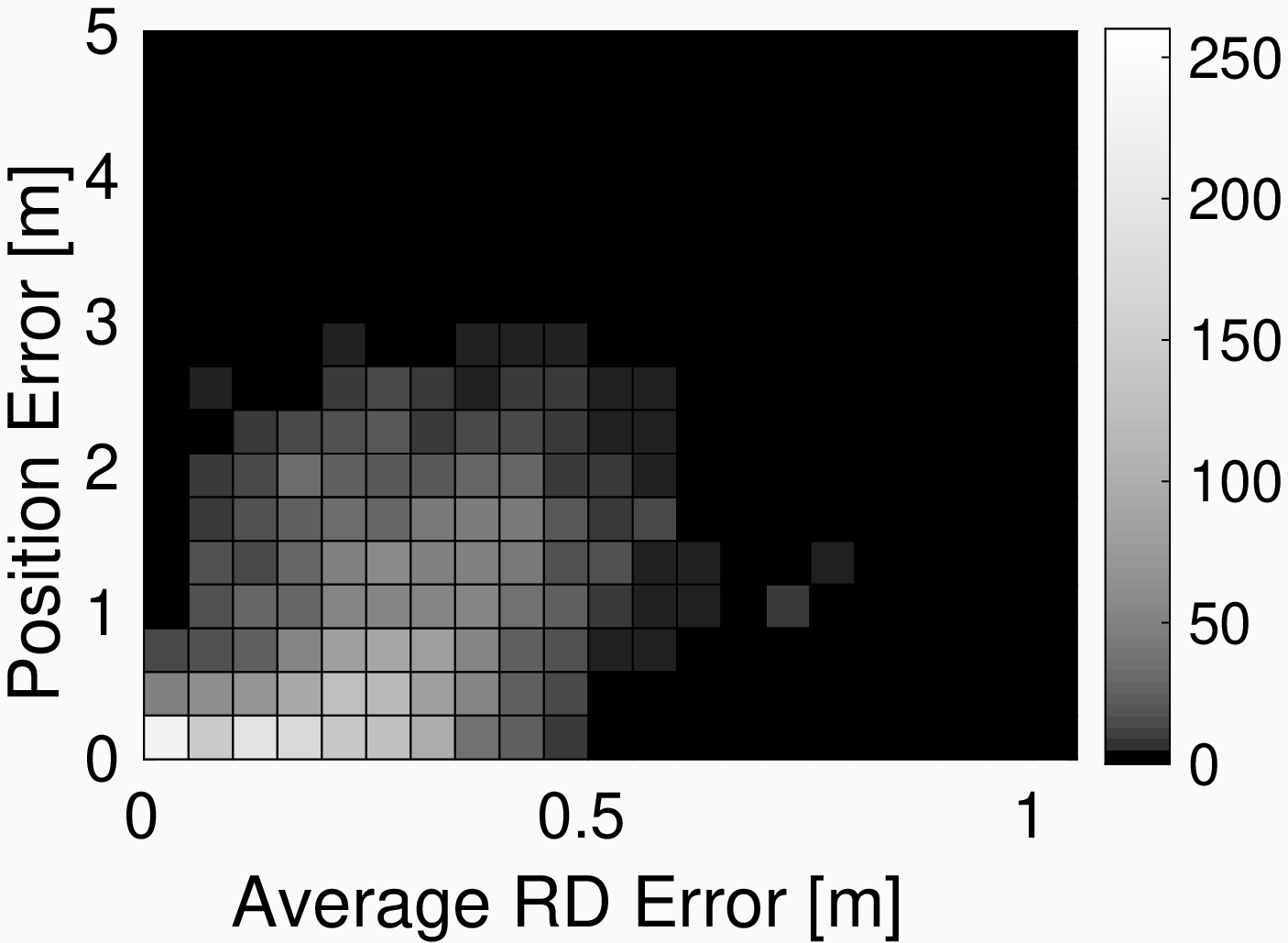}}~\hfill
    \subfloat[usrd-ls]{\includegraphics[width=0.33\textwidth]{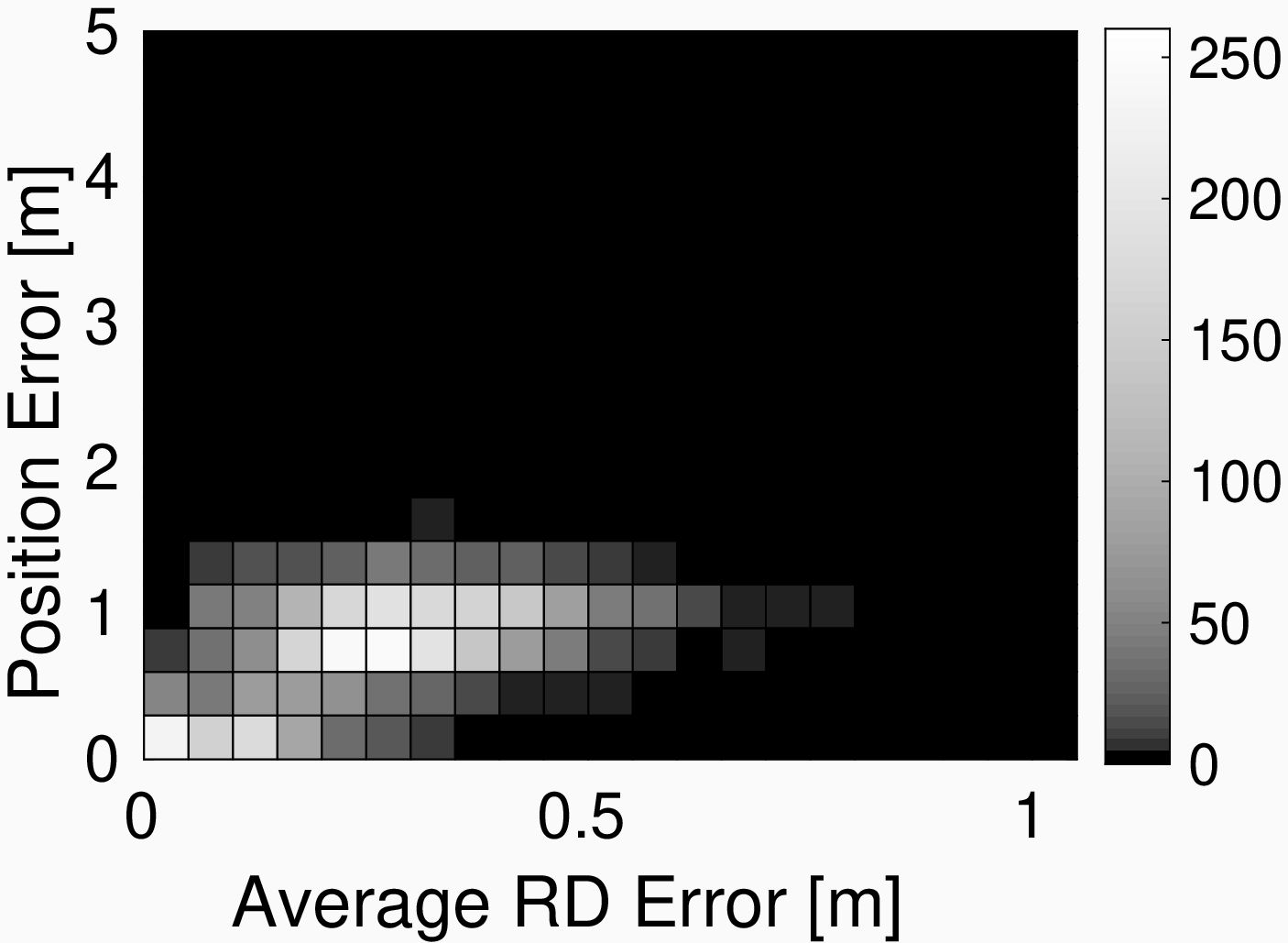}}~\hfill
    \subfloat[srd-ls]{\includegraphics[width=0.33\textwidth]{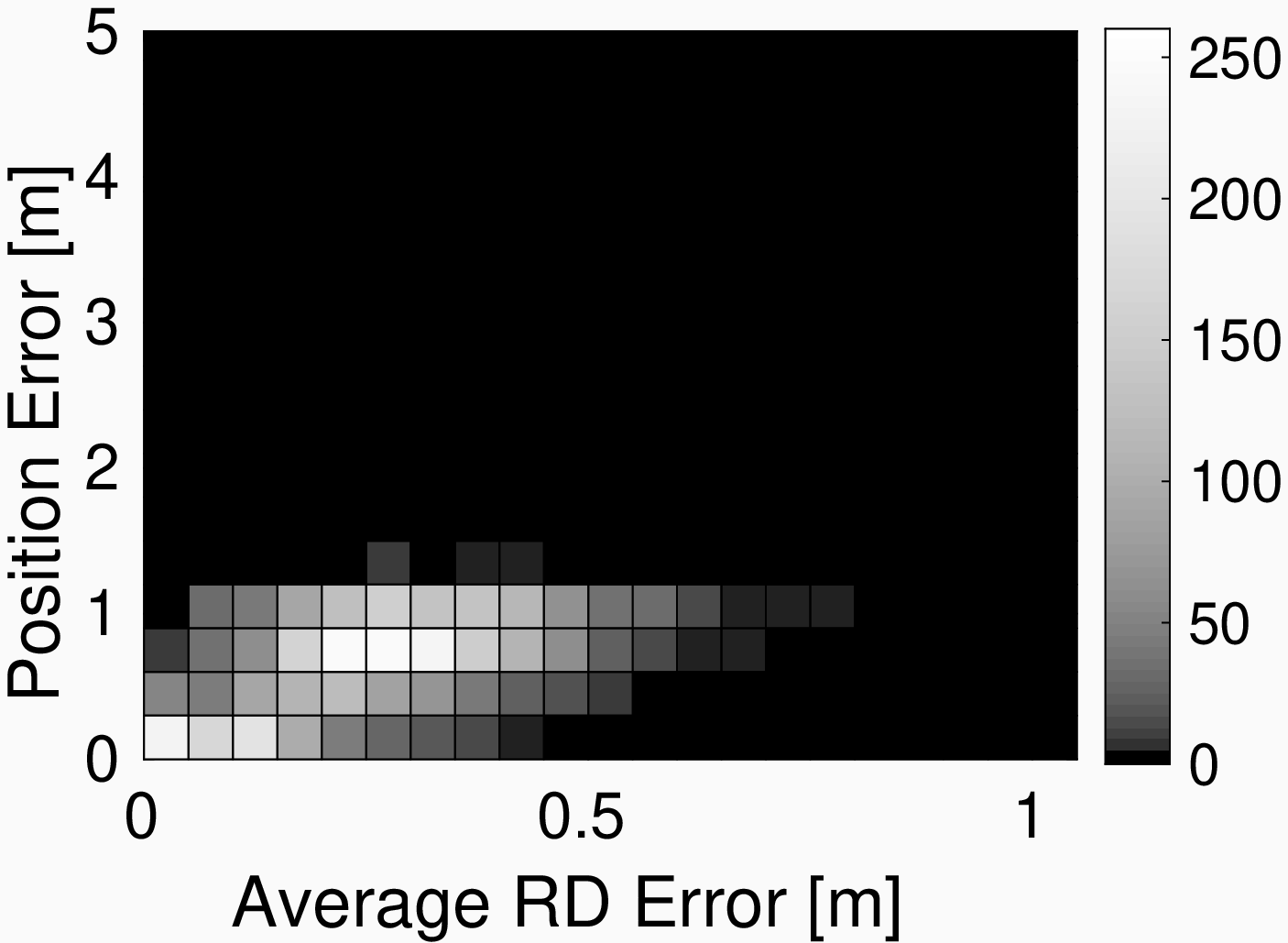}}~\hfill
    \caption{Histogram of the average RD errors \emph{vs} position errors.}
    \label{fig:REvsPosErr}
\end{figure*}

Finally, according to the median multilateration performance plots in the most favorable setting (GCC-PHAT with active VAD and without denoising), the normalized conic LS seems to be the best solution. However, 2D histograms, presented in \ref{fig:REvsPosErr}, which depict the position error with regards to average RD errors, indicate that the constrained spherical LS method offers the most robust performance when the reference microphone is well-chosen (e.g. by selecting the nearest microphone from the barycenter). Related to this choice, a slight drop in performance of both spherical LS methods is observed with the suboptimal (’’min'') choice of the reference microphone.

\section{Conclusion}

Multilateration has a long history, and the methods belonging to this family of localization algorithms are theoretically well-founded. Moreover, they are generic, in the sense that they are essentially agnostic to the signal type, as long as the (pseudo) RDs are obtainable. While this article primarily considers sound source localization, multilateration could be straightforwardly applied to, \emph{e.g.} mapping problems in sensor networks, geolocalization by positioning systems and/or base stations, or to target localization in distributed radar signal processing. 

The multilateration methods of the  ML class are closed to optimal in theory, however they resort to various approximations in order to combat the intrinsic hardness of the localization problem. The LS approaches instead solve easier, but artificial optimization problems. On the other hand, some of them are computationally very efficient, and seemingly work very well in practice. The small-scale benchmark of the three widely used LS methods suggests that constrained spherical LS method offers competitive performance, in terms of accuracy and robustness to TDOA estimation errors, however, at an increased computational cost compared to the unconstrained and conic LS. The choice of the method should be dictated by the use case and the a priori information that may be available: what is the type and the level of measurement noise, how important is the computational complexity, how many microphones comprise the array, what are their specifications etc. There seems to be no clear winner when different criteria are taken into account at the same time, which calls for a dedicated, more comprehensive test study in the future.

\bibliographystyle{IEEEtran}
\bibliography{DistributedLocalization}
\end{document}